\documentclass[pre]{revtex4}

\usepackage{amssymb}
\usepackage{graphicx}

\begin{document}
\title{Forecasting characteristic earthquakes in a minimalist model }
\date{July 30, 2003}
\author{Miguel~V\'azquez-Prada$^{1}$, \'Alvaro~Gonz\'alez$^{2}$,
Javier~B.~G\'omez$^{2}$ and Amalio~F.~Pacheco$^{1}$. }
 \affiliation{$^{1}$Departamento de F\'isica Te\'orica and BIFI. \\
 $^{2}$Departamento de Ciencias de la Tierra.\\ Universidad de Zaragoza. Pedro Cerbuna 12. 50009 Zaragoza. Spain.}

\begin{abstract}

Using error diagrams, we quantify the forecasting of
characteristic-earthquake occurrence in a recently introduced
minimalist model. Initially we connect the earthquake alarm at a
fixed time after the ocurrence of a characteristic event. The
evaluation of this strategy leads to a one-dimensional numerical
exploration of the loss function. This first strategy is then
refined by considering a classification of the seismic cycles of
the model according to the presence, or not, of some factors
related to the seismicity observed in the cycle. These factors,
statistically speaking, enlarge or shorten the length of the
cycles. The independent evaluation of the impact of these factors
in the forecast process leads to two-dimensional numerical
explorations. Finally, and as a third gradual step in the process
of refinement, we combine these factors  leading to a
three-dimensional exploration. The final improvement in the loss
function is about $8.5 \%$.

\end{abstract}

\maketitle

\section{\label{sec:intro} Introduction}

The earthquake process in seismic faults is a very complex natural
phenomenon that present geophysics, in spite of its considerable
efforts, has not yet been able to put into a sound and
satisfactory status. However, in the crucial field of earthquake
prediction, recent years have witnessed significant advances. For
recent thorough reviews dealing with this issue, see Keilis-Borok
(2002); Keilis-Borok and Soloviev (2002), and references therein,
in particular chapter four by Kossobokov and Shebalin. See also
Lomnitz (1994).  The introduction of new concepts coming from
modern statistical physics  seems to add some light and put some
order into the intrinsic complexity of the lithosphere and its
dynamics. Thus, for example,  references to critical phenomena,
dynamical systems, hierarchical systems, fractals, self-organized
criticality and self-organized complexity  are now found very
frequently in geophysical literature (Turcotte, 2000; Sornette,
2000; Gabrielov et al., 1999; Gabrielov et al., 2000). Hopefully,
this conceptual framework will prove its usefulness sooner better
than later.

%Inspired by the sandpile model of self-organized criticality,
We have recently presented a simple statistical model of the
cellular-automaton type which produces an earthquake spectrum
similar to the characteristic earthquake behaviour of some seismic
faults (V\'azquez-Prada et al., 2002). The largest earthquakes on
a fault or fault segment (the events that break its complete
length) are usually termed characteristic (Schwartz and
Coppersmith, 1984; Wesnousky, 1994; Dahmen et al., 1998). For this
reason, in the minimalist model the event of maximum size is
called the characteristic one. Our model is inspired by the
concept of asperity, i.e., by the presence of a particularly
strong element in the system which actually controls its
relaxation. This model presents some notable properties, some of
which will be reviewed in Section \ref{sec:properties}. In Section
\ref{sec:jordan}, an algebraic procedure for the exact calculation
of the probability distribution of the time of return of the
characteristic earthquake is presented. The purpose of this paper
is to quantify the forecasting of the characteristic earthquake
occurrence in this model, using seismicity functions, which are
observable, but not stress functions (Ben-Zion et al., 2003),
which are not. In Section \ref{sec:diagram}, we construct an error
diagram (Molchan, 1997;  Newman and Turcotte, 2002) based on the
time elapsed since the occurrence of the last characteristic
event. This permits a first assessment of the degree of
predictability. In Section \ref{sec:improving}, we propose a
general strategy of classification of the seismic cycles which,
adequately exploited in this model, allows a refinement of the
forecasts. Finally, in Section \ref{sec:discussion} we present
 the conclusions.

\section{ \label{sec:properties} Some Properties of the Model}

In the minimalist  model (V\'azquez-Prada et al., 2002), a
one-dimensional vertical array of length $N$ is considered. The
ordered levels of the array are labelled by an integer index $i$
that runs upwards from $1$ to $N$. This system performs two basic
functions: it is loaded by receiving stress particles in its
various levels and unloaded by emitting groups of particles
through the first level $i = 1$. These emissions that relax the
system are called earthquakes.

 These two functions (loading and unloading) proceed using the following four rules:
\begin{enumerate}
\item In each time unit, one particle arrives at the system.
\item All the positions in the array, from $i=1$ to $i=N$, have
the same probability of receiving  the new particle. When a
position receives a particle we say that it is occupied.
\item If
a new particle comes to a level which is already occupied, this
particle has no effect on the system. Thus, a given position $i$
can only be either non-occupied when no particle has come to it,
or occupied when one or more particles have come to it.
\item The
level $i=1$ is special. When a particle goes to this first
position a relaxation event occurs. Then, if all the successive
levels from $i=1$ up to $i=k$ are occupied, and the position $k+1$
is empty, the effect of the relaxation (or earthquake) is to
unload all the levels from $i=1$ up to $i=k$. Hence, the size of
this relaxation is $k$, and the remaining levels $i>k$ maintain
their occupancy intact.
\end{enumerate}

Therefore, the size of the earthquakes in this model range from 1
up to $N$, being the event of $k=N$ the characteristic one. Note
that the three first rules of this model are exactly those of the
forest-fires model (Drossel and Schwabl, 1992). Our  model has no
parameter and, at a given time, the state of the system is
specified by stating which of the $(i>1)$ $N-1$ ordered levels are
occupied. Each one of these possible occupation states corresponds
to a stable configuration of the system, and therefore the total
number of configurations is $2^{(N-1)}$. These mentioned
$2^{(N-1)}$ stable configurations can be considered as the states
of a finite, irreducible and aperiodic Markov chain with a unique
stationary distribution (Durrett, 1999).

\begin{figure}
 \includegraphics{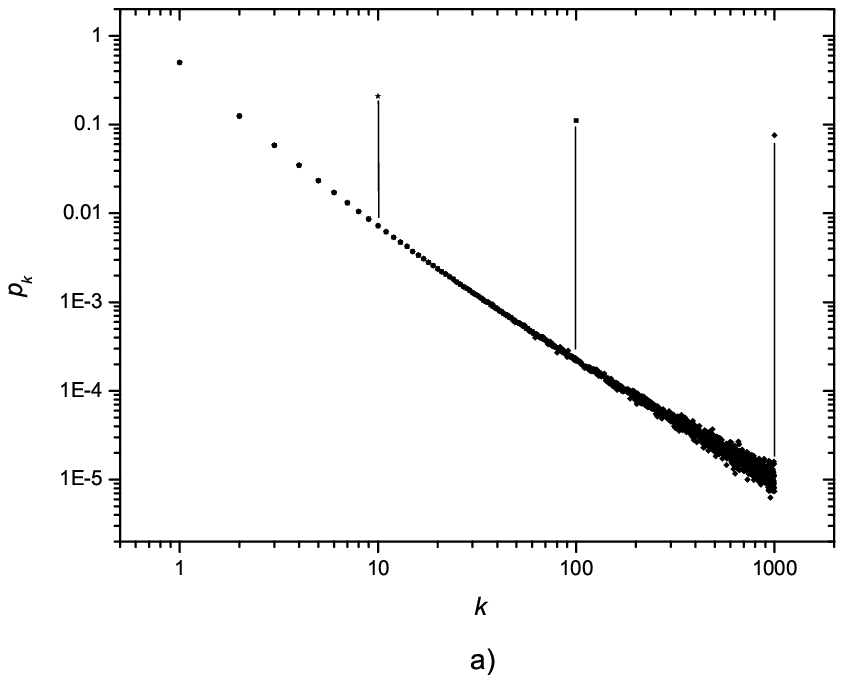}
  \includegraphics{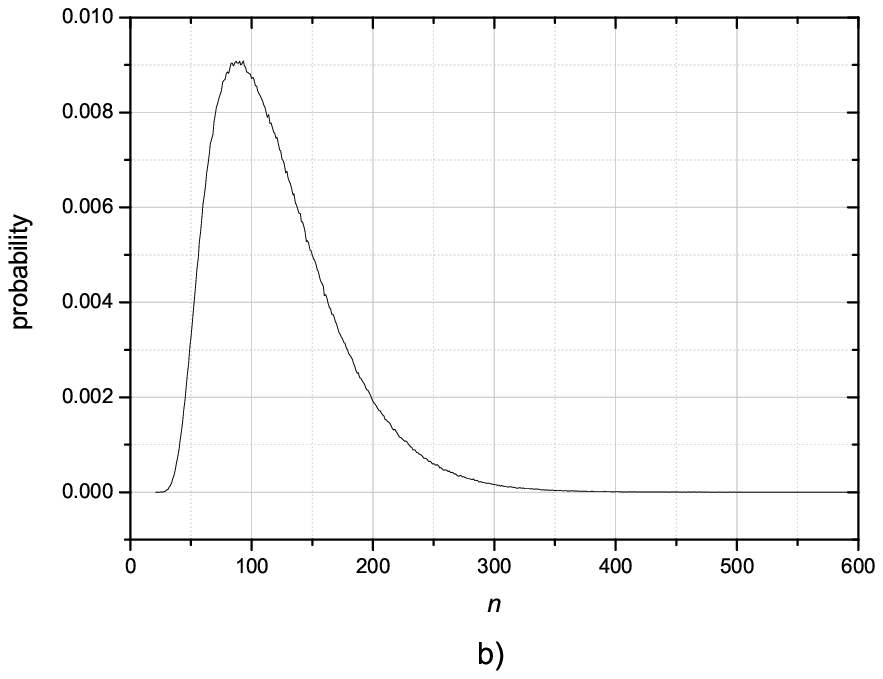}
   \includegraphics{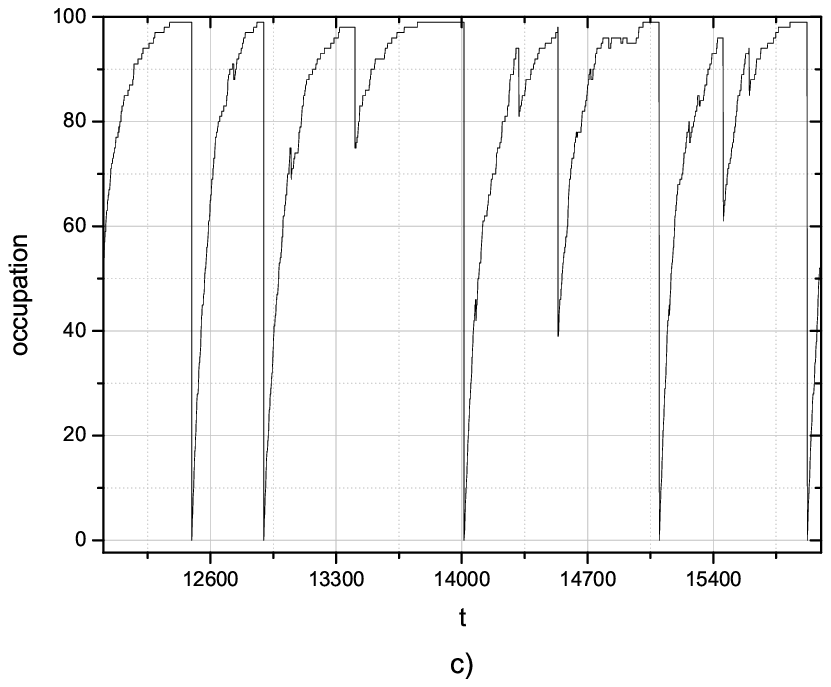}
 \caption{\label{fig:char}
 \ref{fig:char}a) Probability of occurrence of earthquakes of size $k$.
 Note that the simulations corresponding to $N$ equal to 10, 100, and 1000 are superimposed.
 \ref{fig:char}b) For $N = 20$, the probability of return
 of the characteristic earthquake as a function of the time elapsed since
 the last event, $n$.
 \ref{fig:char}c) Time evolution of the state of occupation in a system of
size $N = 100$. Note that after each  characteristic event that
completely depletes the system, there follows the corresponding
recovery up to a high level of occupancy, and then the system
typically presents a plateau previous to the next characteristic
earthquake.}

\end{figure}

The evolution rules of the model produce an earthquake
size-frequency relation, $p_k$, that is shown in
Fig.~\ref{fig:char}a, where the results for $N = 10$, $N = 100$,
and $N = 1000$ are superimposed. Note that this spectrum has a
distribution  of the characteristic-earthquake type: it exhibits a
power-law relationship for small events, an excess of maximal
(characteristic) events, and very few of the intermediate size.
Besides, the three superimposed curves of probability are
coincident.

The result for the probability of return of the characteristic
earthquake, $P(n)$, is shown in Fig.~\ref{fig:char}b for $N=20$.
Here $n$ represents the time elapsed since the last characteristic
event. During an initial time interval $1 \leq n < N$, $P(n)$ is
null, then it grows to a maximum and then finally declines
asymptotically to $0$. (In Section \ref{sec:diagram}, $P(n)$ for
$N=20$, will be usually denoted as curve $a$.) In Section
\ref{sec:jordan}, we explain a general algebraic method for the
exact computation of $P(n)$.

The configurations of the model are classified into groups
according to the number of levels, $j$, that are occupied $(0 \leq
j \leq N-1)$. Using the Markov-chain theory or producing
simulations (V\'azquez-Prada et al., 2002), one easily observes
that in this model the system resides often in the configurations
of maximum occupancy, i. e., in $j = N-2$ and $j = N-1$.

This last property can be observed in Fig.~\ref{fig:char}c, where
we have represented, for $N = 100$,  the time evolution of the
level of occupancy, $j$, in an interval long enough  to observe
the occurrence of several characteristic earthquakes. The typical
pattern after a total depletion is a gradual recovery of $j$ up to
a new  high level of occupancy. Once there, the system typically
presents a plateau before the next characteristic earthquake.
Especially during  the ascending recoveries, the level of
occupancy $j$ suffers small falls corresponding to the occurrence
of rather small earthquakes, that in this model are abundant. Of
course, one also observes that occasionally $j$ falls in a
significant way corresponding to the occurrence of a $N > k \geq
N/2$ intermediate earthquake.

Due to the fact that this model is not critical, it is reasonable
to consider it as an example of self-organized complexity
(Gabrielov et al., 1999).

\section{ \label{sec:jordan} Algebraic Approach to P(n) }

The function $P(n)$, for a minimalist system  of size $N$, is
obtained from the Markov matrix of the system, $\mathbf{M}$,
following the following three steps: i) The element of the last
row and first column of $\mathbf{M}$ is changed by a 0. After this
pruning, the matrix will be called $\mathbf{M}'$. ii) The new
matrix $\mathbf{M}'$ is multiplied by itself $n-1$ times to obtain
$\mathbf{M}'^{(n-1)}$ and the element of the first row, last
column of this matrix is identified. iii) $P(n)$ is the product of
this selected matrix element times $1/N$.

The whys of this recipe are explained in  V\'azquez-Prada et al.
(2002), where $P(n)$ for $N=2$ is explicitly obtained by mere
inspection. The result is
\begin{equation}\label{eq:j1}
P(n)=\frac{n-1}{2^n}, N=2.
\end{equation}
The explicit form of $P(n)$ for larger values of $N$, can be
achieved by exploiting the Jordan decomposition of $\mathbf{M}'$,
\begin{equation}\label{eq:j2}
\mathbf{M}'=\mathbf{Q} \ \mathbf{J} \  \mathbf{Q}^{-1},
\end{equation}
and hence,
\begin{equation}\label{eq:j3}
\mathbf{M}'^{n-1}=\mathbf{Q} \  \mathbf{J}^{n-1} \
\mathbf{Q}^{-1}.
\end{equation}

The matrix $\mathbf{J}$ is formed by ``Jordan blocks'' in the
diagonal positions, i.e., by square matrices  whose elements are
zero except for those on the principal diagonal, which are all
equal, and those on the first superdiagonal, which are equal to
unity. Thus, the task of obtaining an arbitrary power of
$\mathbf{J}$ is simple because, as said, each Jordan block is the
sum of two conmuting matrices: one is a constant times the unity
matrix, and the other is nilpotent. Therefore, in the computation
of any arbitrary power of $\mathbf{J}$, each block is independent
and the corresponding Newton bynomial formula can be applied. As
an example, we now present the calculation of the case $N=3$. In
this case,
\begin{equation}\label{eq:j4}
3  \mathbf{M}'=\left(
\begin{array}{cccc}
1&1&1&0 \\
0&2&0&1 \\
1&0&1&1 \\
0&0&0&2
\end{array}
\right) ,
\end{equation}
which is decomposed as

\begin{equation}\label{eq:j5}
\left(
\begin{array}{cccc}
1&1&1&0 \\
0&2&0&1 \\
1&0&1&1 \\
0&0&0&2
\end{array}
\right) = \left(
\begin{array}{rrrr}
-1&1&1&2 \\
0&0&2&-1 \\
1&1&0&0 \\
0&0&0&2
\end{array}
\right)  \left(
\begin{array}{cccc}
0&0&0&0 \\
0&2&1&0 \\
0&0&2&1 \\
0&0&0&2
\end{array}
\right)  \left(
\begin{array}{rrrr}
-1/2&1/4&1/2&-3/8 \\
1/2&-1/4&1/2&3/8 \\
0&1/2&0&1/4 \\
0&0&0&1/2
\end{array}
\right).
\end{equation}

Therefore,
\begin{equation}\label{eq:j6}
\mathbf{J}^{n-1}= \left(
\begin{array}{rrrr}
0&0&0&0 \\
0&2^{n-1}&(n-1) 2^{n-2}&1/2 (n-1)(n-2) 2^{n-3} \\
0&0&2^{n-1}&(n-1) 2^{n-2} \\
0&0&0&2^{n-1}
\end{array}
\right)  \left( \frac{1}{3} \right)^{n-1} ,
\end{equation}
and from  eq. (\ref{eq:j3})

\begin{equation}\label{eq:j7}
\mathbf{M}'^{n-1}_{1,4} =\frac{2^n}{32} (n-2)(n+5) \left(
\frac{1}{3} \right)^{n-1}.
\end{equation}
Thus, finally,
\begin{equation}\label{eq:j8}
P(n)=\left(\frac{2}{3}\right)^n \frac{(n-2)(n+5)}{32},\ N=3.
\end{equation}

One could optimistically guess that $P(n)$, for an arbitrary $N$,
can be deduced from the systematics observed in the previous
low-$N$ cases. This is disproved by the following formula, which
is the result of $P(n)$ for $N=4$.
\begin{equation}\label{eq:j9}
P(n)=\left(\frac{1}{4}\right)^n \left[ -\frac{13}{16}+\frac{7
n}{4}-\frac{n^2}{2}+\frac{n^3}{32}+3^n  \left[ -\frac{3}{16}+
\frac{7 n}{324}+\frac{n^2}{108}+\frac{n^3}{2599} \right] \right]
,\ N=4.
\end{equation}

Although it is not apparent, this formula, as it should, vanishes
for $n=3$. As in eqs. (\ref{eq:j1}) and (\ref{eq:j8}), $P(n)$ in
eq. (\ref{eq:j9}) is adequately normalized:
\begin{equation}\label{eq:j10}
\sum_{n=N}^\infty P(n)=1.
\end{equation}

\section{ \label{sec:diagram} Error Diagram for the Forecasting of the
Characteristic Earthquake}

In the following paragraphs, we will stick to a model of size
$N=20$ to make the pertinent comparisons. This size is big enough
for our purposes here,  and small enough to obtain good statistics
in the simulations.

For $n=20$ the mean value of $P(n)$ is
\begin{equation}
<n>=\sum_{i=20}^{\infty} P(i)  i = 121.05,
\end{equation}
the standard deviation is
\begin{equation}
\sigma=\left[ \sum_{i=20}^{\infty} P(i)  (i-<n>)^2 \right]^{1/2} =
55.21,
\end{equation}
and the skew of the distribution is
\begin{equation}
\gamma=\frac{1}{\sigma^3}  \sum_{i=20}^{\infty} P(i) \cdot
(i-<n>)^3= -0.10.
\end{equation}

Now we enter into the matter of forecasting. As in any
optimization strategy, we will try to achieve simultaneously the
most in a property called A and the least in a property called B,
these two purposes being contradictory in themselves. Here A is
the (successful) forecast of the characteristic earthquakes
produced in the system. Our desire is to forecast as many as
possible, or ideally, all of them.  B is the total amount of time
that the earthquake alarm is switched on during the forecasting
process. As is obvious, our desire would be that this time were a
minimum. The maximization of A is equivalent to the minimization
of an A$'$ that represents the fraction of unsuccessful forecasts.

Thus, in practice, our goal in this paper is to obtain
simultaneously a minimum value for the two following functions,
$f_e$ and $f_a$. The first represents the fraction of unsuccessful
forecasts, or fraction of failures; the second represents the
fraction of alarm time. These two functions, in this first
one-dimensional strategy of forecasting, are dependent only on the
value of $n$, that is, the time elapsed since the last main event,
and to which the alarm is connected. Using the function $P(n)$
previously defined, they read as follows:

 \begin{equation}
f_e(n) = \sum_{n'=1}^{n}  P(n'),
\end{equation}
\begin{equation}
f_a (n) = \frac{ \sum_{n'=n}^{\infty} P(n')\;
(n'-n)}{\sum_{n'=0}^{\infty} P(n')\: n'}.
\end{equation}

\begin{figure}
 \includegraphics*[width=3.3in]{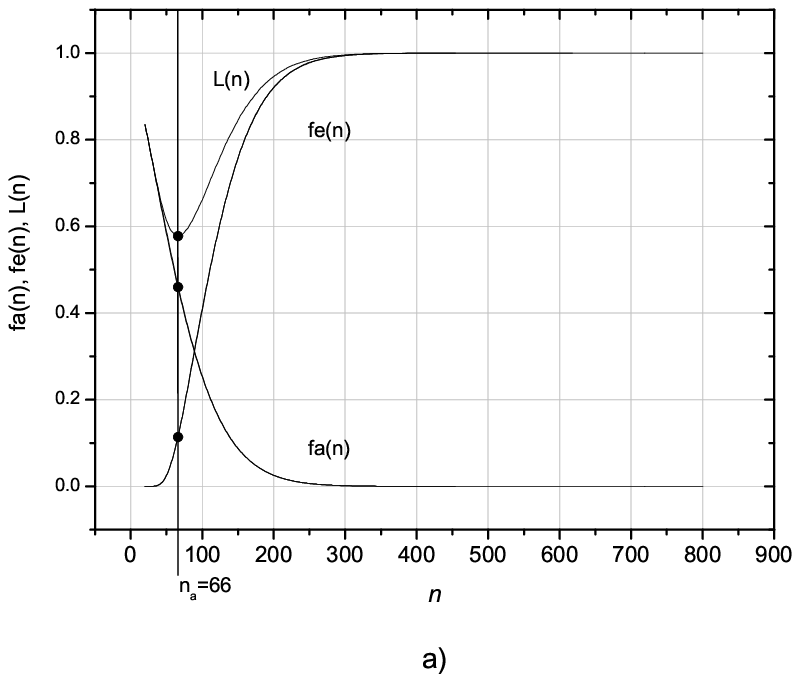}
 \includegraphics*[width=3.3in]{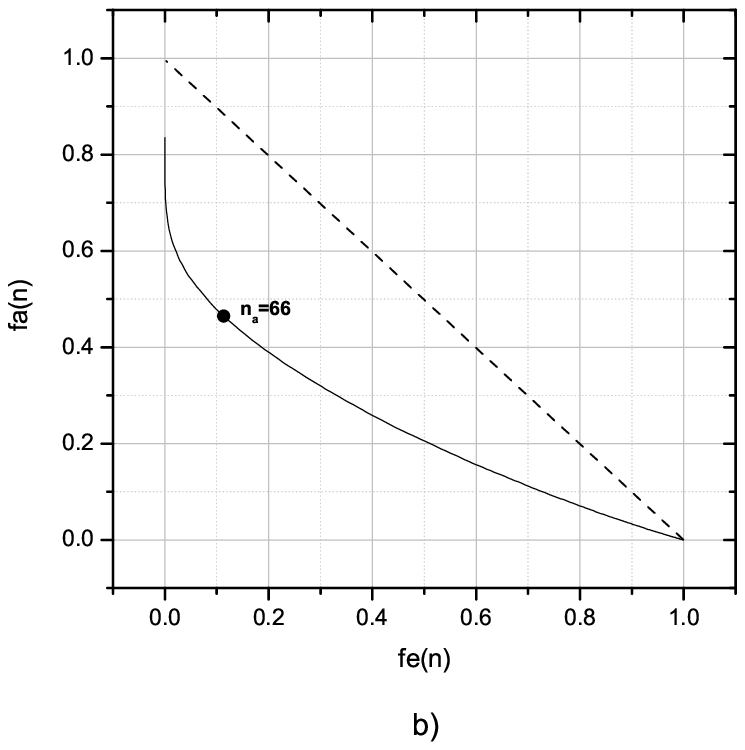}
 \caption{\label{fig:fafe}For $N = 20$, a) Fraction of failures to predict, $f_e$,
fraction of alarm time, $f_a$, and loss function $L=f_a+f_e$ as a
function of $n$. b) Error diagram for characteristic event
forecasts based on $n$. The diagonal line would correspond to a
random strategy.}
\end{figure}

These two functions are plotted in Fig.~\ref{fig:fafe}a. By
eliminating $n$ between $f_e(n)$ and $f_a(n)$, we obtain
Fig.~\ref{fig:fafe}b, which is the standard form of representing
the so-called  error diagram. The diagonal straight line would
represent the result of a random forecasting strategy. The curved
line is the result of this model.

Error diagrams were introduced in earthquake forecasting by
Molchan who contributed with rigorous mathematical analysis to the
optimization of the earthquake prediction strategies (Molchan,
1997). In his papers Molchan used $\tau$ and $n$ to represent the
alarm fraction and the error fraction respectively; and put $\tau$
in the horizontal axis.

To fix ideas, it is convenient to define a so-called loss
function, $L$, which expresses the trade-off between costs and
benefits in the forecasting (Keilis-Borok, 2002). Among all the
possible loss functions, we will choose the simple linear function
\begin{equation}
L=f_a+f_e.
\end{equation}
$L(n)$ is also drawn if Fig.\ref{fig:fafe}a. The position $n_a=66$
provides the minimum value of $L(n)$. $L(n_a)=0.578$. Note that
$n_a$ does not coincide either with  the $n$ that maximizes
$P(n)$, or with $<n>$.

\section{ \label{sec:improving} Improving the Forecasts}

In Section \ref{sec:diagram}  we adopted the strategy of
connecting the alarm at a fixed time, $n$,  after the occurrence
of a characteristic event. The evaluation of this strategy leads
to the conclusion that for $n=n_a=66$, the loss function has a
minimum value $L(n_a)=0.578$. The question now is: Can we think up
other strategies that render better results? To answer this
question, we now return to our previous comments on Fig.
\ref{fig:char}.

If we define a medium-size earthquake as an event with a size
between $N/2$ and $N-1$, i.e. $N > k \geq N/2$, by observing the
graphs in Fig.~\ref{fig:char}, one is led to the conclusion that
in this model the occurrence of a medium-size earthquake is not
frequent but when it actually takes place, the time of return of
the characteristic quake in that cycle is increased.
%(compare, for example, the duration of the second and the third cycles in Fig.
%\ref{fig:char}c).

\begin{figure}
\includegraphics{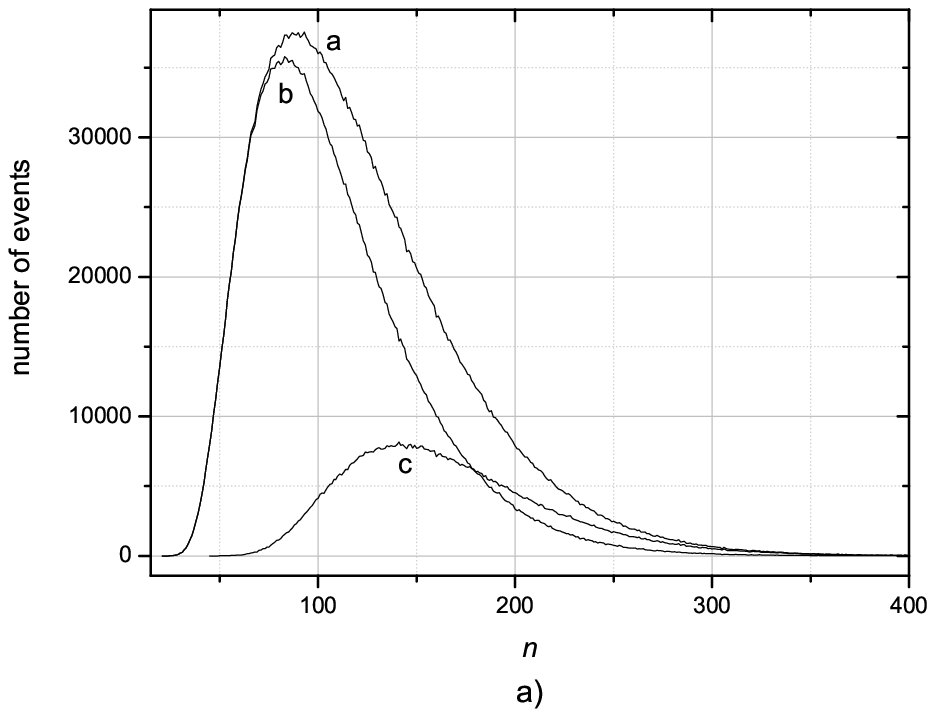}
\includegraphics{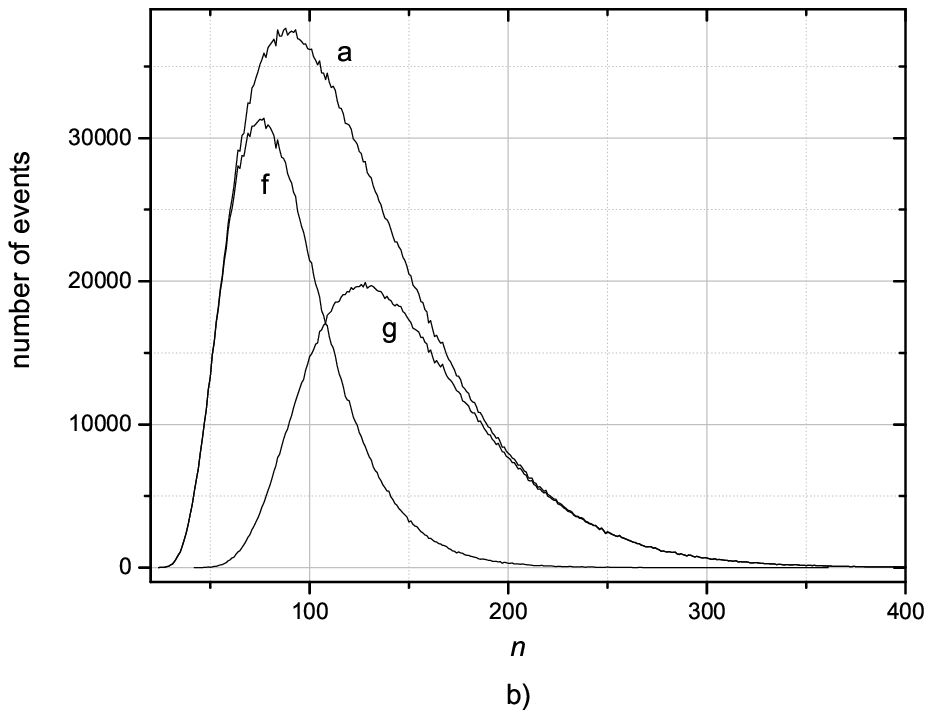}
\caption{\label{fig:split} \ref{fig:split}a) For $N = 20$. Line
$a$ is the distribution of return times of the characteristic
earthquake as a function of the time elapsed since the last event,
$n$. Line $b$ corresponds to the distribution of cycles where no
medium-size earthquake occurs. Line $c$ corresponds to cycles with
medium-size earthquakes. Curves $b$ and $c$ constitute the
splitting of curve $a$ according to whether this retarding effect
is fulfilled or not. \ref{fig:split}b) Lines $f$ and $g$,
represent the separation of the $a$ distribution according to
whether the advancing effect is fulfilled or not.}
\end{figure}

This qualitative perception can be substantiated by numerically
obtaining the probability of having cycles where no medium-size
earthquake occurs, i.e., $k < N/2$. This information is completed
by the distribution of cycles where the condition $N
> k \geq N/2$ does occur. These two distributions are shown in
Fig.~\ref{fig:split}a as lines $b$ and $c$ respectively. Here,
line $a$ represents the total distribution of the times of return
of the characteristic earthquake in this model (the same as
plotted in Fig.~\ref{fig:char}b). Note that, as it should, the
distribution $a$ covers both distributions $b$ and $c$. The mean
time $<n>$ for the three distributions is $<n>_a = 121.05$, $<n>_b
=107.57$ and $<n>_c=166.84$. The fraction of cycles under $b$ is
$0.77$ and the fraction under $c$ is $0.23$. A splitting of this
type, in which the $a$ distribution separates into $b$ and $c$,
will be denoted henceforth as $a=b \oplus c$.

To check if $a=b \oplus c$ is potentially useful for our purposes,
we will now analyze independently these two sets of cycles, $b$
and $c$, with the method used in Section \ref{sec:diagram} for
curve $a$.  The result is the following: the best working $n$ for
dealing with the cycles under distribution $b$ is $n_b= 60$. And
with respect to the cycles belonging to the distribution under
$c$, the best $n$ is $n_c=124$.

\begin{figure}
\includegraphics{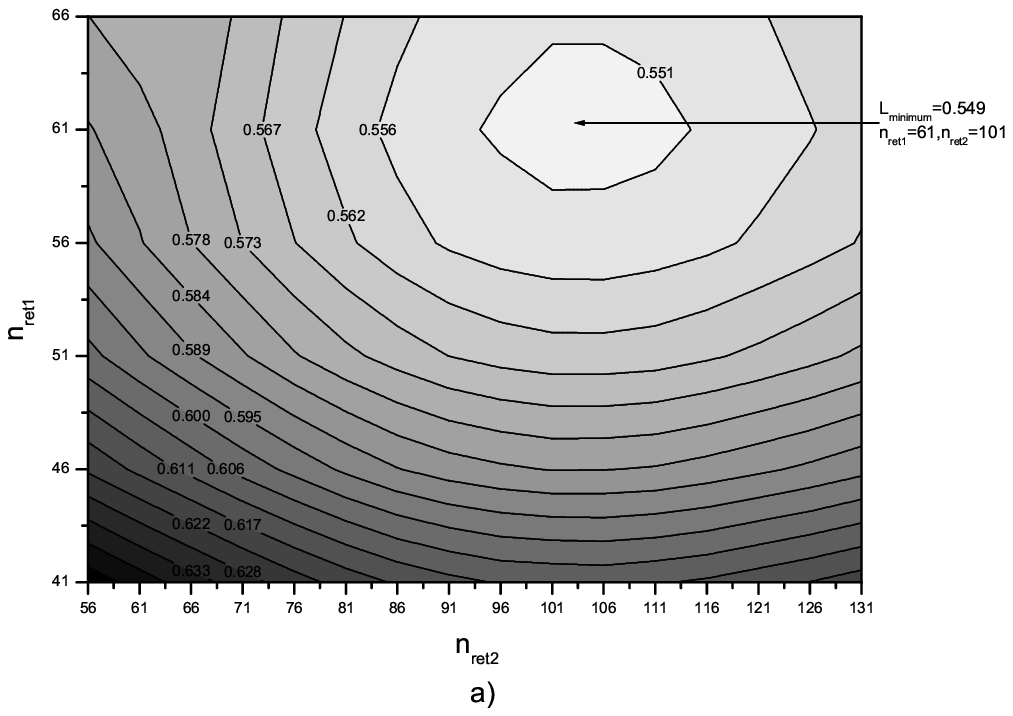}
\includegraphics{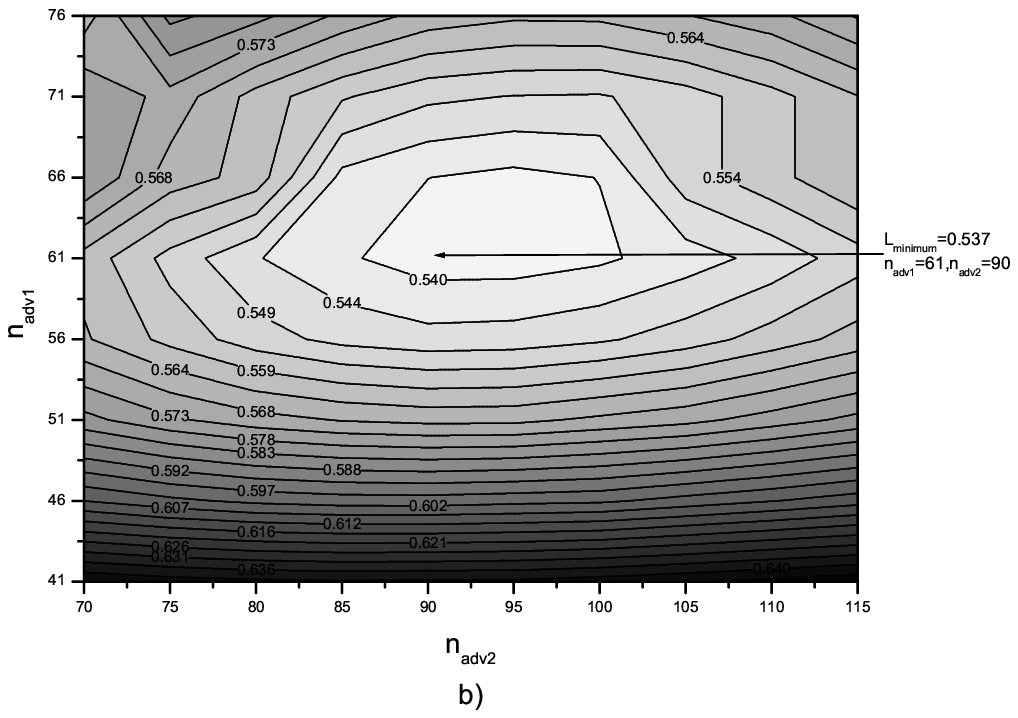}
\caption{\label{fig:2dim} \ref{fig:2dim}a) For $N = 20$. Results
of the two-dimensional strategy based on the splitting of curve
$a$ according to whether the retarding effect is fulfilled or not.
Values of $L$ varying $n_{ret1}$ and $n_{ret2}$.  Minimum value of
$L=0.549$ for $n_{ret1}=61$ and $n_{ret2}=101$. \ref{fig:2dim}b)
Results of the two-dimensional strategy based on the splitting of
curve $a$ according to whether the advancing effect is fulfilled
or not. Values of $L$ varying $n_{adv1}$ and $n_{adv2}$. Minimum
value of $L=0.537$ for $n_{adv1}=61$ and $n_{adv2}=90$.}
\end{figure}

Therefore, we will now study again the whole set of cycles, i.e.
those under $a$, by means of a retarding  strategy, which is based
on the splitting $a=b \oplus c$. We will adopt  the following
steps: in any cycle, we will wait until an $n$, named $n_{ret1}$
(which is near to $n_b$), before taking any decision. If no medium
earthquake has occurred so far, then the alarm is connected at
$n_{ret1}$. If, on the contrary, a medium quake has occurred
before $n_{ret1}$, then we move the alarm to $n_{ret2}$ (which is
close to $n_c$). This notation $n_{ret1}$ and $n_{ret2}$ comes
from the retarding strategy that we are exploring now. This
two-dimensional strategy is implemented by varying $n_{ret1}$ and
$n_{ret2}$ looking for the best value of $L$. This is illustrated
in Fig. \ref{fig:2dim}a. The best option is $n_{ret1}=61$ and
$n_{ret2}=101$, with $L(n_{ret1}, n_{ret2})=0.549$.

Now we look for a similar property that can classify the cycles
from another point of view. This new property consists in
identifying the cycles where the sum of the sizes of all the
earthquakes before the characteristic one is less than $N/2$. This
condition will be represented by $SUM < N/2$.  The reason for this
choice is that if $SUM<N/2$, the system, statistically speaking,
tends to reach more rapidly the configurations of maximum
occupancy, $j=N-2$ and $j=N-1$, and the time of return of the
characteristic quake in that cycle tends to be smaller (see Fig.
\ref{fig:char}c). In Fig.~\ref{fig:split}b, line $a$ represents,
as in Fig.~\ref{fig:split}a, the distribution of return intervals
of the characteristic earthquake for all the cycles of the model.
And lines $f$ and $g$ represent, respectively, the separation of
line $a$ according to the fulfilment, or not, of the $SUM < N/2$
condition, $a=f \oplus g$. The mean value of the $f$ and $g$
distributions is $<n>_f=88.78$ and $<n>_g=151.69$ respectively.
The fraction of events under the $f$ and $g$ lines is $36.96$ and
$63.04$ respectively.

This second splitting of the whole set of cycles in the model,
$a=f \oplus g$, can be used as an advancing strategy in parallel
to what we did with the retarding strategy.
%Underlying this strategy is the well founded suspect that at a time....
%$SUM < N/2$, the system will likely be already in the plateau
%mentioned in Section \ref{sec:properties}, and thus the occurrence
%of a main event is impending.
Thus the independent analysis of curve $f$  leads to  $n_f= 60$,
and the similar analysis of curve $g$ leads to $n_g=90$.

We will now study again the whole set of cycles (under $a$) by
means of the advancing  strategy, which is based on the splitting
$a=f \oplus g$. Therefore, we proceed as follows:  In any cycle,
we wait until  $n_{adv1}$ (which is close to   $n_f$) before
taking any decision. If the condition $SUM < N/2$ has been
fulfilled, then the alarm is connected at $n=n_{adv1}$. If, on the
contrary, this condition has not been fulfilled, then we move the
alarm to $n_{adv2}$ (which is close to $n_g$). This
two-dimensional strategy is implemented by varying $n_{adv1}$ and
$n_{adv2}$ looking for the lowest $L$. This is illustrated in Fig.
\ref{fig:2dim}b. The search for the best option leads to
$n_{adv1}= 61$ and $n_{adv2}=90$, with $L(n_{adv1},
n_{adv2})=0.537$. This value of $L$ is slightly better than that
obtained using the retarding strategy.

\begin{figure}
\includegraphics{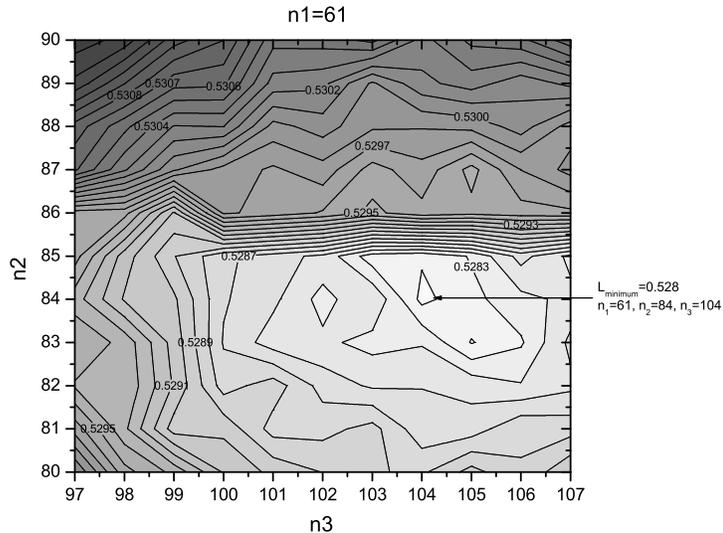}
\caption{\label{fig:3dim} For $N = 20$. Illustration of the
three-dimensional strategy. For $n_1=61$, L-constant level-curves
are plotted. The minimum value of $L$ is $0.528$ for $n_1=61$,
$n_2=84$ and $n_3=104$.}
\end{figure}

Inspired by these results, we will now analize the possibilities
of a mixed strategy which contains conceptual elements of the two
partial strategies discussed so far. Here we will explore a
3-dimensional grid of points $(n_1, n_2,  n_3)$ looking for the
minimization of $L$. The first coordinate, $n_1$,will be explored
in the neighbourhood of $n_{adv1}$,  the second coordinate $n_2$
in the neighbourhood of $n_{adv2}$, and finally $n_3$ near
$n_{ret2}$. The two succesive  key decisions to be taken are:
\begin{enumerate}
\item In any cycle, we wait until $n= n_1$. If  $SUM < N/2$ IS fulfilled, we connect the alarm at  $n_1$ and
leave it there. If at  $n= n_1$, $ SUM < N/2$  is NOT fulfilled,
we move the alarm to $n_2$. And,
\item (We are now at   $n_2$). If no medium-size
event has occurred between $n_1$ and  $n_2$, we leave the alarm
connected at  $n_2$. If, on the contrary, one or more medium-size
events have occurred in this interval, then we move the alarm to
$n_3$.
\end{enumerate}
The search for the triplet  $(n_1, n_2, n_3)$ that makes $L$
minimum  is illustrated in Fig. \ref{fig:3dim}. The result
corresponds to $(n_1=61, n_2=84, n_3=104)$, and there, $L= 0.528$.
This is the best result obtained in this work.

Thus, the  improvement obtained in $L$, when passing from $L(n_a)$
to $L(n1,n2,n3)$ is around $\simeq 8.5 \% $.

\section{ \label{sec:discussion} Conclusions}

In this paper, we have analyzed the behaviour of the  minimalist
model in relation to a quantitative assesment of the forecasting
of its successive characteristic earthquakes. We have chosen a
simple loss function, $L=f_a+f_e$. Our first try, based on a
one-dimensional search in $n$, produces  a minimum result  of $L$
around $0.578$. This was illustrated in Fig. \ref{fig:fafe}a. With
the aim of improving the forecasts, we then explored two modes of
a common strategy  that divides the probability distribution of
the time of return of the characteristic earthquake into two
distinct distributions. The first mode consists in using the
occurrence of intermediate-magnitude earthquakes as a sign that
the characteristic earthquake would likely return at a time later
than usual in that cycle. This is based on the fact that
medium-size events significantly deplete the load in the system
and its recovery induces a retardation. This effect takes place in
any system of the sand-pile type. The exploitation of this idea
leads to a two-dimensional search that finally renders an $L$
value around $0.549$. (Fig. \ref{fig:2dim}a). The second idea
consists in using the fact that a significant absence of small
earthquakes during a sizeable lapse of time in the cycle is a sign
of imminence of the next characteristic event, or at least of a
shortening of its period of return. This strategy is similar to
the old  wisdom in seismology that links a steady absence of
earthquakes in a fault with the increase in the risk of occurrence
of a big  event. The exploration of this idea proceeds similarly
to what we did with the retarding strategy:  this also leads to a
two-dimensional search. It renders a minimum $L$ around $0.537$.
(Fig.\ref{fig:2dim}b).

Finally, a mixed strategy that tries to incorporate the
information acquired is implemented by means of a
three-dimensional search, and provides a value of $L=0.528$. The
identification of the three optimum parameters is illustrated in
Fig. \ref{fig:3dim}.

It is important to remark that the information we have used in our
forecasts is based only in the observed systematics of earthquake
occurrence in the model, i.e., only seismicity functions have been
used. Thus, for example, in Section \ref{sec:improving} we have
not used the state of occupancy of the system $j$, which would
have given much more accurate predictions. In real life, the use
of this information would be equivalent to knowing, in real time,
the value of the stress level and the failure threshold at any
point in a fault.

\subsection*{Acknowledgements}
We are grateful to the two Referees of the first version of this
paper for their thorough and helpful reviews. This work was
supported by the project BFM2002-01798 of the Spanish Ministry of
Science. Miguel V\'azquez-Prada and \'Alvaro Gonz\'alez are
respectively supported by the PhD research grants B037/2001
(funded by the Autonomous Government of Arag\'on and the European
Social Fund) and AP2002-1347 (funded by the Spanish Ministry of
Education).

%\begin{thebibliography}{}\label{sec:biblio}

\section*{ \label{refer} References}
%\bibitem{ben}
\noindent Ben-Zion, Y., Eneva, M., and Liu, Y.: Large Earthquake
Cycles and Intermittent Criticality On Heterogeneous Faults Due To
Evolving Stress And Seismicity, J. Geophys. Res., 108 (B6), 2307,
doi: 10.1029/2002JB002121, 2003. \\
%\bibitem{markov}
Dahmen, K., Ertas, D., and Ben-Zion, Y.: Gutenberg-Richter and
characteristic behaviour in simple mean-field models of
heterogeneous faults, Phys. Rev., E58, 1444-1501, 1998. \\
Drossel, B., and Schwabl, F.: Self-organized critical forest-fire
model, Phys. Rev. Lett., 69, 1629-1632, 1992.\\
Durrett, R.: Essentials of Stochastic Processes, Chapter 1,
Springer, 1999. \\
%\bibitem{gabrielov00}
%
Gabrielov, A., Newman, W. I., and Turcotte, D. L.: Exactly soluble
hierarchical clustering model: inverse cascades, self-similarity
and scaling, Phys. Rev., E60, 5293-5300, 1999.\\
Gabrielov, A.,  Zaliapin,  I.,  Newman, W., and  Keilis-Borok, V.
I.: Colliding cascades model for earthquake prediction, Geophys.
J. Int., 143, 427-437, 2000. \\
%\bibitem{jensen98}
%Jensen, H. J.: Self-organized Criticality.,Cambridge, 1998. \\
%\bibitem{keilis02}
Keilis-Borok, V.: Earthquake Prediction: State-of-the-Art and
Emerging Possibilities, Annu. Rev. Earth Planet. Sci., 30, 1-33,
2002. \\
%\bibitem{keilis02b}
Keilis-Borok, V. I., and Soloviev, A. A. (Eds.): Nonlinear
Dynamics of the Lithosphere and Earthquake Prediction, Springer,
2002. \\
Lomnitz, C.: Fundamentals of Earthquake Prediction, J. Wiley
\& Sons, 1994. \\
%\bibitem{molchan97}
Molchan, G. M.: Earthquake Prediction as a Decision-making
Problem, Pure. Appl. Geophys., 149, 233-247, 1997. \\
%\bibitem{newman02}
Newman W. I., and Turcotte, D. L.: A simple model for the
earthquake cycle combining self-organized complexity with critical
point behavior, Nonlin. Proces. Geophys., 9, 453-461, 2002. \\
%\bibitem{turcotte00}
Sornette, D.: Critical Phenomena in Natural Sciences,
 Springer Verlag, Berlin, Germany, 2000. \\
Schwartz, D.P. and Coppersmith, K. J.: Fault behaviour and
characteristic earthquakes: Examples from the Wasatch and San
Andreas Fault Zones, J. Geophys. Res., 89, 5681-5698, 1984.\\
Turcotte, D. L.: Fractals and Chaos in Geophysics, 2nd Edit.,
Cambridge Univ. Press, 2000. \\
%\bibitem{mvazquez02}
V\'azquez-Prada, M., Gonz\'alez, \'A., G\'omez, J.~B., and
Pacheco, A.~F.: A Minimalist model of characteristic earthquakes,
Nonlin. Proces. Geophys., 9, 513-519, 2002. \\
Wesnousky, S. G.: The Gutenberg-Richter or
Characteristic-Earthquake Distribution, Wich It Is?, Bull. Seism.
Soc. Am., 84, 1940-1959, 1994.

%\end{thebibliography}

\end{document}